\documentclass{svjour3}                     
\smartqed  
\newcommand{\bq}{\begin{equation}}
\newcommand{\eq}{\end{equation}}

\newcommand{\bqa}{\begin{eqnarray}}
\newcommand{\eqa}{\end{eqnarray}}
\newcommand{\fa}{\forall}
\newcommand{\tht}{\theta}
\usepackage{graphicx}
\journalname{Quantum Information Processing}
\begin{document}
\title{Comment on ``Groverian Entanglement Measure and
Evolution of Entanglement in Search Algorithm for $n  (= 3,
5)$-Qubit Systems with Real Coefficients" (Volume 6, Number 4,
August 2007), by Arti Chamoli and C. M. Bhandari}

\author{Preeti Parashar \and Swapan Rana}

\institute{   Physics and Applied Mathematics Unit\\ Indian
Statistical Institute
\\203 B T
              Road\\Kolkata 700 108\\India\\
              \email{parashar@isical.ac.in,
              swapan\_r@isical.ac.in}
           }
\date{Received:   / Accepted: }

\maketitle

\begin{abstract}
We point out that the main results---the analytic expressions for
the Groverian Measure of Entanglement, in the above mentioned
paper are erroneous. The technical mistake of the paper is
discussed. It is shown by an explicit example that the formula for
calculating the Groverian measure yields $G(|\psi\rangle)=0$ for
some entangled states.

\end{abstract}
\section {What is this communication about?}
In their paper ``\textit{Groverian Entanglement Measure and
Evolution of Entanglement in Search Algorithm for $n  (= 3,
5)$-Qubit Systems with Real Coefficients}" \cite{Ch}, Arti Chamoli
and C. M. Bhandari have given an explicit formula for calculating
the `Groverian Measure' of entaglement $G(|\psi\rangle)$ in terms
of the state coefficients. The derivation of the results involves
some maximization process which is very difficult to solve
analytically and we will show that unfortunately the authors have
missed some crucial technical points which has led to incorrect
results.

\subsection {The result gives $G(|\psi\rangle)=0$ for some entangled
states. } Let us consider the $3$-qubit generalized
Greenberger-Horne-Zeilinger ($GGHZ$) state given by
$|\psi_{GGHZ}\rangle=a_{000}|000\rangle+a_{111}|111\rangle,~a_{000}^2+a_{111}^2=1$.
Then using the formula provided in the paper, we get
$P_{max}(|\psi_{GGHZ}\rangle)=1\Rightarrow
G(|\psi_{GGHZ}\rangle)=0$, which contradicts the fact that
$|\psi_{GGHZ}\rangle$ is an (genuinely) entangled state. Indeed it
is well known \cite{Yi} that
\begin{equation}
P_{max}(|\psi_{GGHZ}\rangle)=\mbox{max
}\left\{a_{000}^2,a_{111}^2\right\}.
\end{equation}Similarly, considering the $5$-qubit $GGHZ$ state,
the result leads to a contradiction.

\section {Where has the error occurred?}
The error has occurred at the maximization stage. The maximization
should be with respect to the three independent parameters
$\tht_i,~i=1(1)3$. If one transforms the three parameters to four
parameters, the four parameters cannot be independent and
maximizing the function according to the four parameters
disregards the functional relations between the parameters. Since
$\theta_w,~\theta_x,~\theta_y,~\theta_z$ are not independent, the
maximum of $P(\theta_1,~\theta_2,~\theta_3)$ with respect to
$\theta_1,~\theta_2,~\theta_3$ is in general not equal to the
maximum of $P(\theta_w,~\theta_x,~\theta_y,~\theta_z)$ with
respect to $\theta_w,~\theta_x,~\theta_y,~\theta_z$. The only case
when the two maximums agree is the one when the point of maxima
($\theta_w,~\theta_x,~\theta_y,~\theta_z$) of the four dimensional
space lies on a three-dimensional hyperplane. Since the volume of
this hyper-plane is zero, the probability of randomly choosing
parameters where the optimization with respect to
$\theta_w,~\theta_x,~\theta_y,~\theta_z$ would give correct result
is also zero. Therefore, in the general case, there exists no
$\theta_1,~\theta_2,~\theta_3$ giving
$P_{max}(\theta_w,~\theta_x,~\theta_y,~\theta_z)$.

We will now give an explicit example to elaborate the above discussion. For calculational
simplicity, we will consider the $GHZ$ state.

\subsection {Explicit calculation of $P_{max}$ for $GHZ$ state}
For the state
$|\psi_{GHZ}\rangle=\frac{1}{\sqrt{2}}\left(|000\rangle+|111\rangle\right)$,
\bq
P_{max}(\theta_1,\theta_2,\theta_3)=\max_{\theta_1,\theta_2,\theta_3}\frac{1}{2}
[cos\theta_1cos\theta_2cos\theta_3+sin\theta_1sin\theta_2sin\theta_3]^2\eq
[Note that, from equation (2) it follows by inspection that
$P_{max}=\frac{1}{2}$ and it occurs at
$\tht_i=0\mbox{ or }\tht_i=\pm\frac{\pi}{2}~~\fa~i=1,~2,~3$ ].\\
To get the maximum, just following the mentioned paper \cite{Ch},
we get by converting the trigonometric products into sums,\bqa
P_{max}(\theta_1,\theta_2,\theta_3)&=&\max_{\theta_1,\theta_2,\theta_3}\frac{1}{2}
[cos\theta_1cos\theta_2cos\theta_3+sin\theta_1sin\theta_2sin\theta_3]^2\nonumber\\
&=&\frac{1}{32}\left[(cos\tht_w-sin\tht_w)+(cos\tht_x+sin\tht_x)+(cos\tht_y+sin\tht_y)\nonumber\right.\\
&+&\left.(cos\tht_z-sin\tht_z)\right]^2\eqa where \bqa
\tht_w&=&\theta_1+\theta_2+\theta_3\nonumber\\
\tht_x&=&\theta_1+\theta_2-\theta_3\nonumber\\
\tht_y&=&\theta_1-\theta_2+\theta_3\nonumber\\
\tht_z&=&\theta_1-\theta_2-\theta_3\nonumber \eqa

We emphasize that the maximum in (3) should be taken with respect
to $\tht_1,~\tht_2,~\tht_3$ (and not with respect to
$\theta_w,~\theta_x,~\theta_y,~\theta_z$). This is obtained by
satisfying $\frac{\partial P}{\partial\tht_i}=0$, imposing the
constraint \bqa J_0+J_1+J_2+J_3&=&0\nonumber\\
J_0+J_1-J_2-J_3&=&0\nonumber\\
J_0-J_1+J_2-J_3&=&0\nonumber\eqa
 or equivalently \bq J_0=-J_1=-J_2=J_3\eq where \bqa
 J_0=-sin\tht_w-cos\tht_w&=&-\sqrt{2}cos(\pi/4-\tht_w)\nonumber\\
  J_1=-sin\tht_x+cos\tht_x&=&\sqrt{2}cos(\pi/4+\tht_x)\nonumber\\
   J_2=-sin\tht_y+cos\tht_y&=&\sqrt{2}cos(\pi/4+\tht_y)\nonumber\\
    J_3=-sin\tht_z-cos\tht_z&=&-\sqrt{2}cos(\pi/4-\tht_z)\nonumber\eqa

However, the maximum in (3) with respect to
$\theta_w,~\theta_x,~\theta_y,~\theta_z$ [which has been done by
the authors of \cite{Ch}] is obtained by satisfying
$\frac{\partial P}{\partial\tht_w}=0,\frac{\partial
P}{\partial\tht_x}=0,\frac{\partial
P}{\partial\tht_y}=0,\frac{\partial P}{\partial\tht_z}=0$ imposing
the constraint \bq J_0=J_1=J_2=J_3=0.\eq

Clearly the constraints (4) and (5) are not the same.
\textbf{Indeed, there exists no $\tht_i,~i=1,~2,~3$ in the range
$-\frac{\pi}{2}\le\tht_i\le\frac{\pi}{2}$ which will satisfy the
constraint (5)}.\\

 [Nevertheless, if we use the erroneous constraint
(5) for the maximum with respect to
$\theta_w,~\theta_x,~\theta_y,~\theta_z$, we will get $P_{max}=1$,
a contradiction that $P_{max}$ is the square of the maximum
possible overlap with a fully separable state].

\subsection{Another paper with similar flaw}
We would like to mention that another paper \cite{Ch1} by the same
authors suffers from a similar flaw. In that paper, the authors
have derived an analytic formula to calculate $G(|\psi\rangle)$
for arbitrary 4-qubit pure state $|\psi\rangle$ with real
coefficients. But due to the similar error (as discussed in first
para of Section 2 in the present Comment) in the maximization
process, the formula (in \cite{Ch1}) is incorrect too. Here we
will give just an example to show the discrepancy:

It is well known \cite{Yi,Pr} that for the 4-qubit $W$-state,
$G(|W\rangle)=(\frac{3}{4})^3$. But the formula in \cite{Ch1}
gives $G(|W\rangle)=(\frac{3}{4})^2$.

Thus the formulae presented in the papers \cite{Ch,Ch1} give
incorrect results for both even and odd-qubit cases---as expected.

\section {Conclusion}
Like all other known multi-partite entanglement measures, there is
no explicit expression of Groverian measure in terms of the state
parameters. We need to go through some maximization process which
cannot be achieved analytically for arbitrary parameters. Once the
parameters are specified,  numerical techniques can be used for
maximization. However, for some symmetric states (e.g.
$GGHZ,Balanced \mbox{ and} ~W $ states \cite{Yi}, Dicke states
\cite{Pr}) the Groverian measure can be calculated analytically.

\end{document}